\documentclass[conf]{new-aiaa}
\usepackage[utf8]{inputenc}

\usepackage{graphicx}
\usepackage{amsmath}
\usepackage[version=4]{mhchem}
\usepackage{siunitx}
\usepackage{longtable,tabularx}

\usepackage{url}
\usepackage{CJKutf8}

\title{Education Paradigm Shift To Maintain Human Competitive Advantage Over AI}

\author{Stanislav Selitskiy \footnote{PhD Student, School Of Computer Science And Technology, Park Square, Luton, LU1 3JU, UK} and 
Chihiro Inoue \footnote{Associate Professor in Language Assessment, Centre For Research in English Language Learning And Assessment, Putteridge Bury Campus, Luton, LU2 8LE, UK}}
\affil{University of Bedfordshire, Luton, Bedfordshire, LU1 3JU, UK}

\begin{document}

\maketitle

\begin{abstract}
Discussion about the replacement of intellectual human labour by ``thinking machines'' has been present in the public and expert discourse since the creation of Artificial Intelligence (AI) as an idea and terminology since the middle of the twentieth century. Until recently, it was more of a hypothetical concern. However, in recent years, with the rise of Generative AI, especially Large Language Models (LLM), and particularly with the widespread popularity of the ChatGPT model, that concern became practical. Many domains of human intellectual labour have to adapt to the new AI tools that give humans new functionality and opportunity, but also question the viability and necessity of some human work that used to be considered intellectual yet has now become an easily automatable commodity.
Education, unexpectedly, has now become burdened by an especially crucial role of charting long-range strategies for discovering viable human skills that would guarantee their place in the world of the ubiquitous use of AI in the intellectual sphere. We highlight weaknesses of the current AI and, especially, of its LLM-based core, show that root causes of LLMs' weaknesses are unfixable by the current technologies, and propose directions in the constructivist paradigm for the changes in Education that ensure long-term advantages of humans over AI tools.
\end{abstract}

\section{Introduction}
\lettrine{A}{n} explosive debut in public of the ChatGPT \cite{BibEntry2023Mar2} and the following similar Large Language Models (LLM) \cite{BibEntry2023Mar4, chowdhery2022palm, BibEntry2023Mar3, touvron2023llama} also initiated a debate on LLMs' effects on education. An obvious first reaction was concern about abusing the LLMs' ability to generate human-like texts for cheating and plagiarism \cite{orenstrakh2023detecting} in such examinations and tests that evaluate students in such faculties as memorisation, summarisation, reviewing, and basic analysis. Various methods of detection and prevention of using LLMs in education and academia were proposed \cite{tang2023science,khalil2023will,rodriguez2022cross,savelka2023can}.

However, the next wave of publications on the place of LLMs in education started to contemplate the thought that even if education shut the doors before LLMs, the industry would not, such as putting graduates who are not accustomed to the use of LLMs at a disadvantage. The publications started coming to the conclusion that education itself should change, not pursuing obsolete goals and not executing obsolete practices \cite{anders2023using,rudolph2023chatgpt}, but instead concentrating more on the areas where human-lead education (even armed with LLMs as tools) has advantages over mere LLMs in themselves \cite{fuchs2023exploring,cope2019education}.

The impact of AI on education and academia became even a global concern, and UNESCO (The United Nations Educational, Scientific and Cultural Organization), to help meet this challenge, published a guidance \textit{Guidance for generative AI in education and research} available in six languages \cite{unesco2023}. Although it is a high-level overview document, it is a good start for thinking and discussion. Among 
several key points, UNESCO's guidelines underscores the importance of the human-centred approach and fostering critical thinking, in that AI in education should prioritize human development and be used to support and enhance the educational experience. Suggestions include the use of AI for handling routine tasks and providing personalized learning paths and support as a digital teaching assistant. This would allow educators more time for interactions in the classroom, such as group activities, discussions, and one-on-one mentoring sessions, to foster students’ critical thinking, emotional intelligence, and social interactive skills. UNESCO states that ensuring that AI supports, rather than diminishes, these interactions is key to maintaining a balanced educational environment. Such use of AI will force educational institutions to make changes not only to their curriculum, but also to their syllabi, lesson contents, grading systems, as well as what skills and competencies they aim to nurture in students. 

Reconsidering the target skills and competencies that educators and students need in the era of generative AI demands careful reviews of what LLMs do and do not do well. While discussions of LLM and other Generative AI abilities and their impact on all aspects of human culture are somewhat shallow, a deeper and more important question of how relatively simple algorithms lay in the foundations of LLMs, when scaled up, resulted in impressive human-like text generation. The answer to this question may have a very heavy impact on our understanding of what natural human intelligence and language are, and maybe they are not as glorious and unique phenomena as we used to think. In the same way, on top of the obvious implications of the LLM and other Generative AI use and misuse in education, the deeper questions are related to the common or dominant education paradigms, problems of which were highlighted and surfaced up by AI. On these deeper aspects of education in the new AI era is what we want to focus in our research.

The collection of the LLMs' weaknesses, without a detailed survey of their strengths and beneficial uses, may seem like a ``sour grapes'' complaints. However, obvious utilitarian benefits from LLM use in the wide but still limited applications are out of the scope of this paper, We are concerned with LLM weaknesses exactly for the purpose stated in the title - to find gaps in LLM functionality which still could be applications of human strength, and to channel education in the direction of developing these traits and skills.

The contribution is organised in the following manner: Section~\ref{llm_observ} reviews the research in the area of empirical testing of the LLM abilities, Section~\ref{llm_flaw} lists design principles LLM architecture are built on and their inherent limitations, Section~\ref{ling_prob} shows linguistic views on the LLM organisation and function, Section~\ref{trans_prob} lists problems with Transformer architecture of artificial neural networks LLMs use, Section~\ref{llm_cant} lists week LLM points and suggests humans to concentrate on them, Section~\ref{ed_change} proposes constructivist paradigms of education which are better suited to prepare humans to rivalry with LLM/AI, Section~\ref{ed_current} lists gaps in the current education practices that should be bridged to move to the constructivist paradigm, Section~\ref{ed_method} proposes methodological instruments that may help to convert current education practices to the envisioned, Section~\ref{studen_ed_method} proposes application methodological instruments to implement desired education paradigm, and Section~\ref{conclude} draws final conclusions.

\section{Empirical Observations of the LLM's weaknesses}
\label{llm_observ} 

From the literary text analysis perspective, the generated by LLMs, though usually syntactically correct, are effete, emotionless washed-up texts, lacking linguistic variability and distinctness, and pragmatic intercity and originality \cite{gao2022comparing,chaves2021impact,wilkenfeld2022ai,mitrovic2023chatgpt}. On the dynamic debating or deliberation text generation, LLMs also perform far from ideal. For example, on detecting discourse move, ChatGPT performed even worse than simple BERT models \cite{deliang2023}. Debates with ChatGPT, as everybody can see using the OpenAI interface, suffer from circular arguments, self-contradiction, and evasiveness - tendencies to please human preferences in Reinforcement Learning (RL) \cite{ramamurthy2022reinforcement,carta2023grounding} - exactly those practices that nobody wants to foster in students. When used to detect manipulative discussion tactics of cyberattacks, ChatGPT also scored significantly worse than simple BERT models \cite{fayyazi2023uses}.

General LLMs' problems with functional domains such as mathematics, reasoning, and logic \cite{frieder2023mathematical}, emotional expressivity, wit, humour and ethics \cite{borji2023categorical,Arkoudas2023Jan}, factual data, privacy, and false, bias and discrimination \cite{Basta2019, Kurita2019, Sheng2019, Gehman2020, BibEntry2022Sep, Bianchi2022, Weidinger2021, tang2023does, Goldstein2023} are well documented. 
Machine Learning (ML) specific problems of LLMs add such issues as lack of interpretability and understanding, \cite{bender-koller-2020-climbing, Lake2020, Marcus2022, Ouyang2022, Leivada2022, Ruis2022}, and catastrophic ageing and forgetting by LLMs \cite{lazaridou2021pitfalls, Amba_Hombaiah_2021, Dhingra_2022, MCCLOSKEY1989109, PARISI201954, ratcliff1990connectionist, kirkpatrick2017overcoming, huang2023survey}. 
LLMs lack agency, structural representation of the language, and real-world picture \cite{Browning2022Aug,Floridi2023}.

Experiments with detecting the use of LLMs and comparing LLM-generated texts with human-generated produced mixed results \cite{gao2022comparing,casal2023can}. 

However, LLM detection and LLM-generated text analysis experiments were conducted on short and static texts, such as article abstracts or isolated fragments of conversation with LLM. Research of a dialogue with LLM as a whole is not numerous, but very fascinating. The ``chain-of-..." family of methods grounded on gradual conditioning nudging of LLMs into the desired direction \cite{wei2022chain}. It was proven to be a perspective practical approach, on which Retrieval Augmented Generation (RAG) industry techniques are based on \cite{lewis2020retrieval}. However, a more interesting direction is mutual human-LLM manipulation \cite{scheurer2023technical,park2023ai}, especially demonstrating the deceptive behaviour of LLMs even without being directly instructed to do so, but indirectly pursued. A similar power of persuasion is observed in the lying games between LLM-driven agents \cite{o2023hoodwinked}. Research on lying and deception generation and its subsequent detection demonstrates a shallow horizon of LLM lie cover-up \cite{pacchiardi2023catch,hagendorff2023deception}. However, if LLMs are explicitly trained to cover up lies and deception, it may be difficult to provide safety measures against such a behaviour \cite{hubinger2024sleeper}.

In our study \cite{Selitskiy2024LLM}, we thoroughly document long conversations with representative evasive moves and turns of ChatGPT. Although the study discusses complicated grammatical aspects of the Japanese language in detail, ChatGPT stumbled upon and had difficulties handling and recovering from erroneous conclusions. The main point of the study is not to point out these errors but instead to attract attention to pathological behavioural patterns of ChatGPT in a dialogue, which is unacceptable for a well-structured, high-quality human discussion.

Despite the inevitable errors, which are a natural part of discussions and debates, even among humans, it's apparent that ChatGPT exhibits some of the most egregious rhetorical fallacies commonly seen in human conversation, which are deemed unacceptable in civilized discourse. These flawed rhetorical practices include evasive verbosity, attempts to anticipate and fulfil the interlocutor's desires, inconsistency and lack of integrity in arguments, and opportunistic changes in stance. These observations resonate with prior studies \cite{scheurer2023technical,park2023ai}, which demonstrate that subjecting models to narrative pressures, such as anticipated financial or political repercussions, can induce deceptive behaviour in LLMs. In our case, the anti-chain-of-thought pressure is more subtle, nevertheless reasonably effective - the pressure of doubt.

Another similarity with existing research on deception is that LLM behaves without conviction as a ``leaky bucket'' without explicit training for counter-deception detection resilience, changing position easily \cite{pacchiardi2023catch,hagendorff2023deception}. And not once but multiple times, repeatedly contradicting itself. That is also observed in \cite{tyen2024llms}, that LLMs can correct themselves but can not detect when they are in error. Where is, and whether it exists at all, the final true point of LLM conjectures, original, or being helped by the ``chain-of-...'' family of prompt engineering methods? Human feedback converges LLMs to their biases (sycophancy) instead of truth \cite{sharma2023towards}. In the same way we pushed ChatGPT towards our desirable subjective opinion equilibrium, could we keep pushing it towards other subjective opinions in the demonstrated ``jittery mode'' of operation?

\section{Fundamental Foundations of the LLMs' Flaws}
\label{llm_flaw}

Although implementation details of the latest models are kept proprietary, previously published research shows that LLM models are built and trained using three main principles. Traditional Natural Language Processing (NLP) tokenizing techniques include the preprocessing stage, on which ``stop-words'' are removed, remaining words are stemmed and lemmatized (converted to canonical dictionary form), and the Bag of Words (BoW) algorithm is used to map lemmatized words into a linear vector space, spanned on the most frequent and important words dictionary basis. The whole sentence or a bigger text is represented as a linear sum of all token vectors (or also so-called ``embeddings'') \cite{Zhang2010}. Such an approach is very resource usage effective but does not count in the sentence or larger text structure. For example, such sentences as: ''A dog bites a man'', ``A man bites a dog'', and ``Dogs bite men'' would be represented by the same embedding.

To introduce implicit elements of the linguistic structures, modern NLP models frequently use context tokenizers \cite{taylor1953cloze} of the BERT-like family \cite{devlin2018bert}. A simple illustration of the BoW and BERT embedding differences would be the former creating ``DOG'', ``BITE'', ``MAN'', and the latter - ``nullDOGbite'', ``dogBITEman'', ``biteMANnull'', ``nullMANbite'', ``manBITEdog'', ``biteDOGnull''. That solves the BoW's structure blindness problem but greatly increases the dimensionality of the embedding space, which is the starting point of LLMs' high computational demands and size.

The second foundation technology the LLMs use is based on the statistical n-gram approach \cite{brown1992class}. The supervised training of the Machine Learning (ML) models has a bottleneck in the manual labelling of the training data sets. To process high amounts of text and other media, LLM uses a self-supervised approach based on the Masked Language Model (MLM) \cite{salazar2019masked,besag1975statistical}. In such a paradigm, part of the words are kept hidden from the ML model in training, and the purpose of the training is to find words with the highest probability of being in the hidden positions. Again, such an approach does not directly model linguistic structures but implicitly stochastically takes them into account.

To keep with the human reader's attention span and produce a coherent flow of text, LLMs have to use long context windows for MLM training of thousands of words. The brute force use of the whole continuous windows is computationally problematic. Therefore another technique of extracting the most valuable and influential context words on the predicted word gave birth to computationally tractable but still huge LLMs - Attention mechanism \cite{bahdanau2014neural, luong2015effective, gehring2016convolutional} and its Transformer implementation \cite{VaswaniSPUJGKP17}. In such an approach of ``self-attention'', learnable matrices are used to compute cosine or Euclidean distances between the word relevance to the projected prediction over the context window sliding, and the most consistent contributor over time is kept and used, in such a way reducing computational demand.

The stochastic nature of the LLMs in modelling structured natural languages has been a point of fierce debate since the LLMs introduction \cite{Bender2021, Schick2020, Marcus2018, Blodgett2021, Bommasani2021}.

Another obvious problem of LLMs is the naivety of their language representation from the theoretical linguistics perspective that operates with categories of syntactic and semantic structures. The former are various kinds or relations in the mathematical sense \cite{combe2022geometry,marcolli2023mathematical}, specific to particular languages, which endow non-ordered multi-sets of the morphing lexemes and are continuously mapped to the universal semantic structures (of meaning or of thought) \cite{chomsky2023genuine} (or, possibly, to universal grammar) \cite{watumull2020rethinking}. Building models of such complex relations in LLMs, capable of discovering and retrieving such linguistic structures and, in such a way, achieving explainability and interoperability of LLMs, is a drastically undeveloped area of research \cite{deletang2022neural}, frequently limited to naive methods of asking LLMs about their internals \cite{jiang2020can}.

\section{Fundamental Linguistic Problems of LLMs}
\label{ling_prob}

The innate non-sequential, hierarchical, and non-local nature of the natural human languages \cite{lyons1968introduction} causes difficulties for the predominately consecutive LLM algorithms. The terminology used in this problem formulation is borrowed from the stratificational view of grammar \cite{lamb1964sememic}. Although, in other branches of cognitive linguistics or other linguistic schools, terminology may vary but express the same idea \cite{watumull2020rethinking}. 
The natural human language can be viewed as layered or stratified (rooted into neuro-cognition mechanisms \cite{lamb2016linguistic}), for example, phonetic, lexical, syntactic, and semantic. Elementary units of one layer, such as lexons (stems, suffixes, prefixes), build composite units, such as lexemes (words), which on the next strata serve as elementary units, such as morphons to be composed into morphemes, building syntax of the sentences, and then semems building semantics (meaning) of the text.

Noam Chomsky, who created a whole new branch in cognitive linguistics, especially emphasises the non-locality of such synthetic units. In inflectional languages such as Balto-Slavic or agglutinating such as Japanese, the non-locality is obvious because of their free word order, but even for the significantly sequential analytic English, Chomsky referees at the semantic attachment of an adverb to a correct verb regardless of their position and order, for example in ``Intuitively, birds that fly swim'' \cite{berwick2016only}. Chomsky proposes a nested binary set concept for biologically plausible complex structures of the natural language. Such sets would allow the merging of sequentially distant lexons into arbitrary complex lexemes, morphemes, and sememes \cite{berwick2011biolinguistic}.  

LLMs are largely ``black box'' models, prone to adversarial attacks, unexpected and strange for humans \cite{zou2023universal}. LLMs' mechanisms introduce implicit naive syntax emulation elements by projecting hierarchical tree structures on flat sequences but with the loss of complexity. For example, in Chomsky's example, ``Intuitevely'' can become the sequential neighbour of ``swim'' by dropping ``fly''.
We hypothesise that the current limited LLM functionality \cite{deletang2022neural}, based on the poor representation of the complex hierarchical syntactic relations of the natural human languages, can be improved only if their more sophisticated, linguistics-informed modelling, which requires a significant breakthrough in the current technologies. 

The explainability and interpretability of LLMs is an underdeveloped area of research, mostly concentrating on answering the question ``How LLMs do it?'' by analysing weights of BERT tokenisers \cite{wu2020perturbed,vig2020bertology} and activations of Transformers \cite{alammar2021ecco}, or n-gram/MLM probabilities \cite{katz1987estimation,lavrenko2017relevance}. As for the questions ``What LLMs do?'' and ``What is the meaning of that?'', some researchers, considering the probabilistic nature of LLMs, think these questions meaningless \cite{Bender2021, Marcus2018, Bommasani2021}. While others wait for LLMs ``emerging abilities'' \cite{kosinski2023theory,bubeck2023sparks} to ask LLMs themselves \cite{jiang2020can}. Again, the former group is quite sceptical about these ``abilities'' \cite{ullman2023large,sap2022neural}.

From the linguistics view on natural human languages, universal semantic roles and relations between parts of a sentence, for example ``Elmer threw a porcupine to Hortense'', such as Actor (Elmer), Patient (porcupine), and Beneficiary (Hortense) could be mapped to syntactic roles and relations, specific to particular languages \cite{marantz1981nature}. In English, syntactic relations between Subject, Direct and Indirect Objects are marked by the order and prepositions (to); in languages such as Balto-Slavic - by the case (nominative, accusative, dative) suffixes; in Japanese - by particles (\begin{CJK}{UTF8}{min}を, に\end{CJK}).

However, the question of what is the language of semantics/meaning, or the ``language of thought'', and how it is externalised into syntactic structures, is difficult even for linguistics and neuroscience of the natural human languages \cite{gallistel2011prelinguistic}. Nevertheless, our hypothesis is that the poor performance of the LLMs in reflective and emotional expressivity is rooted in the lack of semantic structure modelling, and introducing explicit learning of these structures will improve such expressivity.

\section{Fundamental Technical Problems of Transformers}
\label{trans_prob}

Transformer architecture for artificial neural networks (ANN), especially for the encoder-decoder (or just encoder) models, gained tremendous popularity with the publication \cite{VaswaniSPUJGKP17} in which previous works on the attention mechanisms \cite{bahdanau2014neural,gehring2016convolutional,luong2015effective} were compiled and repackaged into a multi-head model which was dubbed as ``Transformer'', and applied to the Natural Language Processing (NLP), and in particular Machine Translation (MT) task.

The attention mechanisms, particularly the dot-product proximity mechanism of Vaswani's Transformer, help to solve the bottleneck's inadequate dimensionality problem of auto-encoders. The too-narrow bottleneck could lose important parameters, while too-wide would ineffectively use computational resources for processing low-useful parameters.

For NLP tasks, such parameters could be sequences of words or, rather, their embeddings into the contextual token (feature) space. In other application domains, such as image \cite{dosovitskiy2020image}, video \cite{liu2022video}, time series processing \cite{verma2021audio}, or even graphs \cite{min2022transformer}, Transformer architecture can also be beneficial, finding important for the task-related parameters, such as images, pixels, temporal signals, or relations.

However, despite their popularity, Transformer architectures exhibit a number of problems of various kinds; some of them are effectively solved in a practical sense, and some are open discussion topics, such as their poor generalization under the Out-of-Distribution (OOD) conditions \cite{yadlowsky2023pretraining}, catastrophic loss of dimensionality, i.e. degradation to a rank-1 matrix over multiple layers   \cite{dong2021attention}, loss of plasticity and forgetting \cite{pelosin2022towards,shang2023incrementer}, to be fair, that latter one is a problem in general for Deep Learning (DL) architectures.

The deepest fundamental flaw of the Transformer architecture is another side of its strength - it scales up or down amplitude either of the whole observation depending on the cosine proximity to majority of other observations in the batch or their linear transformations or separate components of the observations (depending on the variations of the architecture) \cite{Selitskiy2023Batch}, which makes the architecture blind to rare and atypical observation or more complex than linear relations between the observations.

\section{Summary of What LLMs Can't Do Very Well, and Humans Better Do It Better}
\label{llm_cant}

The crucial question of whether LLMs can model thought and intelligence, although receiving a number of optimistic answers \cite{kosinski2023theory,bubeck2023sparks}, still is answered negatively by many \cite{ullman2023large,sap2022neural}. Surprisingly, in the last years, the voices of the critics of the limitations of the traditional narrow ML (and LLMs as part of it), such as Noam Chomsky and Garry Marcus, were joined by such big names of the narrow ML as Joshua Bengio \cite{LexClips2023Aug}, Yann LeCun \cite{BibEntry2023Aug}, and even Geoffrey Hinton whose students built ChatGPT \cite{Metz2023May}.

Still, despite the voiced doubts, the current LLM and Generative AI paradigm stays unchanged because it has not yet been exhausted in the practical sense. Therefore, its fundamental flaws will stay for quite some time. We will list them here in a concise form so human advantages over them can also be listed, and ways of developing them can be formulated. 

\begin{itemize}
  \item LLMs are incapable of creating an integral model of the real world.
  \item LLMs don't have agency (unless viewed from an indirect Latourian \cite{Latour2007} sense) and active pursuit of knowledge and understanding.
  \item LLMs don't have personal positions, and act to please the user. They are ``stochastic parrots'' imitating the trivial ``competent mediocrity'' of the Internet.
  \item LLMs don't have a concept of the other, and possible cooperation in society (of LLMs).
  \item LLMs lack a personal model of genuine emotional expressivity. Generic imitation of it feels fake.
  \item LLMs learn data as a whole, and if forced into incremental learning, they forget previously learned and eventually lose the ability to learn new data.
  \item LLMs can't generalise complex non-sequential, hierarchically or net-structured information - only particular examples of such structures mapped to sequences.
  \item LLMs are tremendously computational, energy, and ecological resources hungry.
\end{itemize}

Now, when we listed LLMs' weak points, we can envision areas where humans may perform superior, employing their abilities, to LLMs and Generative AI. Currently, those abilities can be in low demand, and even not welcomed by society. But if we want to establish and maintain human superiority over LLMs, society should put forth and support those abilities and practices.

\begin{itemize}
  \item Ability to build complex and non-contradictory world models. That may seem like a vague, meaningless lip service to humans, but the immediate consequences of such a maxima are quite radical. Humans need leisure time and material resources to build such models. Compulsory indoctrination by the established religious, social, and economic worldviews damages quality world models. 
  \item Ability and encouragement to safely exercise own initiative in the lifestyle, communication with other humans and states, ways of acquiring knowledge.
  \item Personal opinion should be protected from the pier, professional, social or state pressure in the nether form. In short, ``academic freedom'' should be extended to the whole society.
  \item The cooperative nature of human beings in any interaction should be respected not only in terms of demands and obligations but, in the first place, in terms of rights, rewards, and status.
  \item Fake and cliche communication are markers of behaviour unsuitable for humans.
  \item Continuous lifetime learning is desirable, highly valued, and rewarded human ability.
  \item Universality and wide spectrum of interests, the resurrection of the ``Renaissance Person'' contributing in many areas.
  \item Previously considered unaffordable in the mass society education practices, compared to LLMs' costs become affordable.
\end{itemize}

Such non-commodified abilities to behave not like LLM (LLMs behaviour is described by Ben Goertzel as ``competent mediocrity'' \cite{Charrington2023Apr}), will remain in high value and demand. We want students to be ``competent'', for which goal LLMs may be useful tools and positive examples, but also not to be ``mediocre'', for which LLMs also may be used as counter-examples in their problematic areas. 

\section{How Education Can Foster Human Advantages Over AI}
\label{ed_change}

The currently widely adopted approach to education envisions knowledge transmission from the teacher to the student, from the position of authority of the former. Jean Piaget challenged this paradigm from the child development psychology positions, and then developed further by Lev Vygotsky. From a methodological point of view of general knowledge acquisition, this approach was extended by Georgy Shchedrovitsky \cite{vygotsky2012thought,beilin1992piaget,Shchedrovitsky1995}.

From the constructivist point of view of Piaget-Vygotsky on student psychology of education, teaching-learning is not a forced process of knowledge transfer, but instead, the construction of knowledge about the world, which can be helped, directed, and shaped by the instructor, but fundamentally is guided by the student initiative and developing capabilities which grow alongside the very knowledge acquisition in the settings of social interactions.

The basis for such an approach has formed in rejecting both Pavlovian naturalistic reflexes-based explanations of the thought process, and Freudian mentalist explanations of thought process by other thought processes. In Piaget-Vygotsky's view, the thought process required a methodology of its study outside of it, based on actions caused by the thought process and affecting it. In Shchedrovitsky's terminology, we can only study the compound Thought-Action phenomena.

Apparently, if the thinking-learning-teaching process is indivisible from actions, such actions are performed inside the society and using tools, though specific ``psychological'' tools, to build or construct students' knowledge about the world. If for Piaget such a construction was more student-driven, for Vygotsky, it was more a cooperative effort of both student and teacher whose efforts meet in the ``zone of proximal development'' (ZPD).

Such an approach to education sets significantly higher standards for teachers/instructors, where there is no need for merely transmitting texts, examinations targeted at memorisation, and standardized assessment metrics. In such an absence of formalized metrics and the introduction of inevitable subjectivity into students' assessments, higher ethical standards may be demanded from the teachers. In addition, the potential need for multiple teachers per student was considered prohibitively costly. 


\section{Is Traditional Education Paradigm Up to the Task of Fostering Human Advantages Over AI?}
\label{ed_current}

The constructivist paradigm for education envisions practical implementation of methods of education of constructing knowledge and understanding, such as (but not limited to):

\begin{itemize}
  \item Fostering a big picture view, understanding, and based on them, first-hand actionable application, experimentation and implementation of the knowledge.
  \item Continuous, recursive (i.e. changing assignments) feedback (aizuchi - a rare Japanese loan into English linguistic jargon \cite{kita2007nodding}).
  \item Pursuit of student questions and interests. Interactive (i.e. self-assigning) and co-acting (together with pedagogue) learning.
  \item Non-disciplinary or non-didactic learning, self-involved assessment.
  \item Dynamic knowledge acquisition, with each step in it being a challenge for the student, seemingly impossible, but with guidance and work achievable, building confidence in own abilities.
  \item Collaborative, social learning - learning through cooperation and teaching other students.
  \item Emotion and sentiment expression aware and competent learning and teaching.
\end{itemize}

How realistic is such an approach in the context of the recent educational trends, which are based on mandatory attendance, standardized curriculum and testing, continuous quizzing, multiple-choice, and closed-book exams? Unlike high-cost constructivist approach which advocates for subjective assessment and therefore requires highly skilled teachers, current education trends has allowed for a drastic reduction in education cost, reduction of teacher responsibility and, hence, reduction of qualification, low per-student resource usage ratio, increased access to such education, and obtaining easily-calculated efficiency metrics for better accountability. 

However, despite appearances of higher efficiency, equality and democratisation of such trends in education, deep-level reasons and meaning of such a paradigm were challenged in terms of their origins and goals. The origins of the high-stakes standardised testing are related to eugenics \cite{au2013hiding}, and sadly they are still used to maintain inequality and market-driven education financing and control \cite{rear2019one}, limit freedom and creativity of both teachers and students and enforces goals alien to the nature of education \cite{riffert2005use}.

Fortunately or unfortunately, with contemporary LLMs easily passing standardized tests in many areas \cite{de2023can,maitland2024can,newton2023chatgpt}, and such making the ``teaching to the test'' education irrelevant, this discussion also becomes meaningless - neither such an education paradigm nor students prepared by it are no longer needed in the current world of Generative AI.

Despite the ubiquity of the standardized curriculum and testing paradigm, Wilhelm von Humboldt's program is still considered a theoretical foundation of contemporary Western-type universities. Although its principles, when looked at closely, may seem radical in today's pragmatical reality. For example, Humboldt's principles include \cite{gunther1988profiles,scott2022invoking}:
\begin{itemize}
  \item Freedom in teaching, learning, and research - professors have an unalienable right to choose what they wish to research and teach.
  \item Nobody has a monopoly on truth. Not even the most renowned scientists, and even less, no administrators, politicians, or the public.
  \item Civilized State (or other centres of our in nowadays world) that respects the above freedoms and doesn't interfere with them, under no disguise.
  \item Students' self-cultivation, self-formation, self-understanding - developing their unique full potential and growing into a person only they can become.
  \item Students' participation in the selection of professors and administration.
  \item Unity of teaching-learning, research and knowledge.
  \item Students' attendance is optional, but participation in research is mandatory. It is students who do research, and professors only guide and assist.
  \item Students create new demands in the course of the learning and research process.
\end{itemize}

Humboldt's principles of valuing student development and autonomy, as well as encouraging research to build knowledge rather than simply transmitting it have much in common with Piaget-Vygotsky's constructivist education approach and today's UNESCO's guidelines. However, with such a radical discrepancy between education ideals and practices of the commercialized universities as "successful failing institutions" \cite{scott2022invoking}, how could they be reasonably reconciled?

\section{Transition of the Current Education Practices into the Forms Targeting Human Advantages Over AI?}
\label{ed_method}

Obviously, stating that theoretical ideals of the Humboldtian education system parted ways with their practices is not enough to fix the problem of a large part of human intellectual work becoming obsolete in the era of LLMs and Generative AI. However, works of the Moscow Methodological Circle, which built a lot on Vygotsky's ideas, and which is represented by the mentioned above Georgy Shchedrovitsky \cite{Shchedrovitsky1995}, as well as Vladimir Lefebvre \cite{Lefebvre2010}, Eric Yudin \cite{Blauberg1977}. Ideas of the system-structural methodology of thought-action, including such group activities, and methods of the cross-disciplinary integration and synthesis of the group Thought-Action, named ``activity-organisational games'' can give us tools not only for the synthesis of the student-teacher group activity synthesis but also for the synthesis and reconciliation of the all involved parties in the new AI-oriented change of educational paradigm.

\begin{figure}
\includegraphics[width=0.9\linewidth]{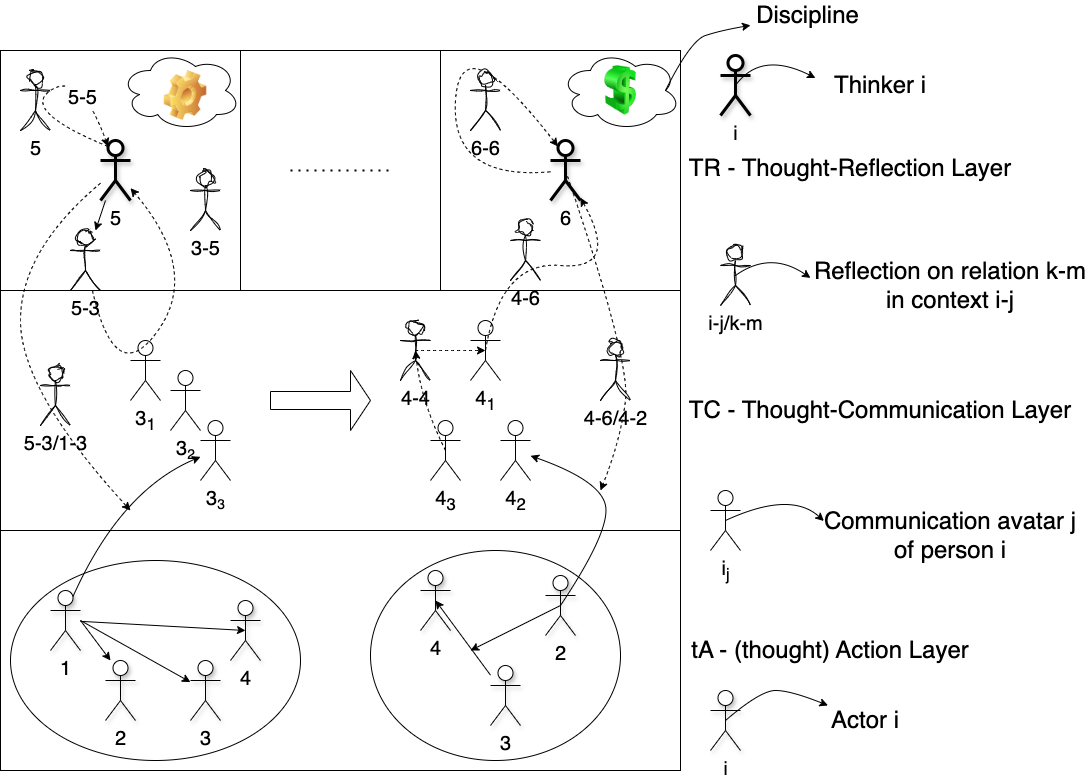}
\caption{Thought-Action concept layers: Thought-Reflection, Thought-Communication, and (thought)Action.\label{fig.ta}}
\end{figure}   

The concept of Thought-Action (TA), simplified elements of which can be found in the well-known Observe-Reflect-Plan-Act loop, consists of three layers: Thought-Reflection, Thought-Communication, and (thought)Action, Figure~\ref{fig.ta}. 
\begin{itemize}
\item The first layer of Thought-Reflection (TR) represents individual non-verbal thinking processes of individuals, including logical constructs, insights, opinions on others, and self-reflection. 
\item The second layer of Thought-Communication (TC) is an inter-person verbal or non-verbal layer of exchanging opinions, disagreements, and suggestions in a natural language form or other communication forms as, for example, graphics. Texts of this layer can not be wrong or right - they are just a means of transferring opinions.
\item The third layer of (thought)Action (tA) is a layer of a socially, culturally, or politically structured 
 group or collective actions. 
\end{itemize}

In a healthy state, the whole thought-action system integrates all three layers, which work together in consort in a continuous interaction manner. However, pathological changes may isolate some or all layers, which inevitably leads to their degradation, crisis, and destruction.

For example, in the chapter \textit{The methodological meaning of the opposition of the naturalistic and the activity-system approaches} \cite{Shchedrovitsky1995}, Schedrovitsky describes the isolated second Thought-Communication layer detached from other layers as follows: ``TC can eliminate its reflection connection with tA and TR and develop immanently only on the limits of TC reality, turning into actionless and meaningless speech, into a pure play of words, without organising and providing neither for TR, nor tA''. Any who is involved in LLM research or even uses it can find these words very familiar.

Or, tA, left isolated from TR and TC layers, ``therefore becoming stagnant mechanical self-reproduction, devoid of life and all mechanisms of meaningful change and development''. The latter quotation describes nicely the situation with an educational approach that stresses high-stakes formalised standardized closed-book tests, ignoring its eugenic origins and radical departure from the Humboldtian education principles. For example, in \cite{topirceanu2017breaking} a creative and cooperative behaviour of students combating standardized closed-book tests \cite{lucifora2015cheating}, which is the goal for Humboldtian University and Piaget-Vygotsky education, is called ``dishonest'' and ``cheating'', while proposing surveillance on students' social media accounts with the aims of detecting friendship and collaborative connection, and breaking them to prevent cooperation. The unethical or even questionable legality of such activities, more typical for scammers, and lost high-level aims of the educator are not reflected upon.

Instruments of the ``activity-organisational games'' proposed by the Methodological Circle are aimed at employing the three-level Thought-Action concept for integrating diverse individuals from various disciplines, patterns of thinking, operations and world views for solving unprecedented new complex tasks, creating novel protocols and methods by constant communication between participants, individual reflection on how personal patterns needs of thinking and behaviour needs to be changed to understand and cooperate with others, and immediate iterative correction and modification of actions in the group based on the communication and reflection going in the Observe-Reflect-Plan-Act loop.
Modifying the education model to target human competitiveness over AI will likely require transforming the deepest beliefs about what education is about, what kind of tests and examinations are needed, or needed at all, and what constitutes plagiarism, dishonesty, or cheating.

\section{Methodological Approach to Implement Piaget-Vygotsky's Constructivist Education and Reanimate Humboldtian University Principles}
\label{studen_ed_method}

In the absence, or at least severe limitation, of the Humpboltdian education principles of freedom of learning and teaching, student-educator cooperation and mutual knowledge construction goals, and predominance of the paradigm of the knowledge transfer from the higher authority position, student and educator opposition and misunderstanding is inevitable. Hence, the use of the ``activity-organisational games'' to reconcile the Thought-Action of both students and educators would also be beneficial. 

The above-mentioned ethically questionable educator attitude \cite{topirceanu2017breaking} is not a single occurrence. In \cite{goerisch2024considering}, authors argue that the introduction of digital surveillance of attendance, plagiarism checks, invasive online exam proctoring, and recorded Zoom sessions, swiftly implemented during the COVID-19 crisis, and framed by the universities as acts of care, in reality, creates an atmosphere of distrust and harm.
On the other hand, unproctored high-stakes closed-book exams lead to inflated scores \cite{carstairs2009internet}. The attitude of the university faculty, in \cite{bujaki2019utilizing}, was found to be similar to the financial institutions' toward financial fraud, and therefore concentrated more on the ``opportunity'' factor of the well-accepted in the financial domain ``fraud triangle'', rather than to rethinking ''pressure'' and ``rationalization'' factors. For example, instead of ramping up invasive surveillance during closed-book exams, switch to open-book exams in which the use of external materials is not considered ``cheating''. Especially in the light of research demonstrating the advantages of open-book exams in advanced subjects \cite{damania2021remote,ramamurthy2016study,malone2021effect,theophilides1996major,williams2009efficacy}. Or, from students' perspective, would be much better to replace exams completely with more meaningful, research-related activities, such as paper reviews \cite{sletten2021rethinking} or research portfolios \cite{vigeant2021portfolio}. Of course, educators, university faculty and administration need to communicate with students, listen to them, rethink their attitudes and implement changes \cite{sonbuchner2022reconnecting,cacciamani2012influence}, for which ``activity-organisational games'' proposed by the Methodological Circle is a very useful instrument.

Another stage of such mutual adaptation of students and educators can be at the individual level of the ``design your exam'' approach when a particular student and professor can discuss what forms of knowledge construction program and verification are more suited for their individual case \cite{shahba2021design}. While the form of the examination itself is not necessarily is ``good'' or ``bad'', it is about how outcomes are used \cite{ragusa2017s}. Some students may be better at memorisation, and some may be better at understanding. Wrong examination delivery may be corrected by the educator, who may give a high passing mark to a student who formally failed all standard choice questions but, upon further examination, expressed a deep understanding of the topic. Or, \textit{vice versa}, when a formally correct student has no real understanding of the matter. 

However, such an approach requires great freedom and trust in the educators' qualifications and intentions. This raises questions about the subjectiveness, compatibility, and social aspects of student-teacher personal and social relations. When a student ``cheats'' during an exam or brings an LLM-generated essay, or the teacher is hostile and unfair to a student, it means that they have just not arrived at Vygotsky's ZPD. Whose fault is it? Maybe both, but what is more important, it is the failure of both... and that is perfectly normal. In the Piaget-Vygotsky paradigm, failure is always an option. Cheating, freedom and trust, and as a result, subjectivity, are social concepts \cite{alan2020cheating} and need social cures when failure of a particular student-teacher pair is not the high-stake one, and do not lead to catastrophic or even significant financial or career consequences.

The need for dedicated, skilled educators under the Piaget-Vygotsky and Humboldtian models calls not just for a drastic increase in their compensation and social status but also for a widening recruitment basis, especially from industry practitioners who have troves of experience and may be eager to change their career path. That is one reason for the increase in the cost of future, AI-aware education. But we can not afford not to afford such changes if we want to keep human education relevant in the ubiquitous AI era.

\section{Conclusions}
\label{conclude}

Education as a knowledge transmission paradigm is no longer adequate for the world where AI takes routine, former ``intelligent'' labour from humans. Such an educational approach stresses developing skills in students, which puts humans at a disadvantage to AI. However, AI of the current development paradigms has significant flaws, leaving competitive niches for humans. Current AI acts as ``(superficially) competent mediocrity'' which is ``frequently wrong, but never in doubt'' and suffers from ``hallucinations'' and lack of trust.

However, despite all of these problems and the ecological dangers of noosphere pollution by mediocrity (in V. Vernadsky's sense), LLMs successfully beat humans in standardized tests in many domains and other exams requiring simple memorisation.
This success of LLMs ends the long discussion about the effectiveness and relevance to real education of standardized tests, closed-book exams, and cost-effective, lowly-paideducator staff. It no longer matters if such an education approach correlates to the actual success of students in academic and industrial positions.

Therefore, to foster the needed skills in students, new education paradigms are needed, as UNESCO's guidelines suggests. In this article, we focused on Piaget-Vygotsky's constructivist approach to education and Humboldtian concept of University as a place for much more flexible collaboration between students and educators. Previously, such education paradigms were considered unaffordably costly. However, humans can no longer afford not to change education principles in the wake of AI.

Such a radical educational paradigm shift requires no less radical instruments for reconciling the highly diverse views of various stakeholders, including teaching and research academia, educational administration, students, industry, and policy-makers. As such an instrument, ideas of the Methodological School on Thought-Action are proposed, as well as the instrument of flexible and individually tailored implementation of the constructivist approach in the education system.

\bibliography{ref}

\end{document}